\documentclass[aps,prl,onecolumn, preprint,superscriptaddress,showpacs]{revtex4}

\usepackage{graphicx}
\usepackage{bm}

\begin{document}
\title{Laser-driven collimated tens-GeV monoenergetic protons from mass-limited target plus preformed channel}

\author{F. L. Zheng}
\affiliation{Key Laboratory of HEDP of the Ministry of Education,
CAPT,and State Key Laboratory of Nuclear Physics and Technology,
Peking University, Beijing, China, 100871}\affiliation{Graduate
School, China Academy of Engineering Physics, Beijing, China,
100088}

\author{S. Z. Wu}
\affiliation{Key Laboratory of HEDP of the Ministry of Education,
CAPT,and State Key Laboratory of Nuclear Physics and Technology,
Peking University, Beijing, China, 100871}\affiliation{Institute
of Applied Physics and Computational Mathematics, Beijing, China,
100088}

\author{H. C. Wu}
\affiliation{Key Laboratory of HEDP of the Ministry of Education,
CAPT,and State Key Laboratory of Nuclear Physics and Technology,
Peking University, Beijing, China, 100871}

\author{H. B. Cai}
\affiliation{Key Laboratory of HEDP of the Ministry of Education,
CAPT,and State Key Laboratory of Nuclear Physics and Technology,
Peking University, Beijing, China, 100871}\affiliation{Institute
of Applied Physics and Computational Mathematics, Beijing, China,
100088}

\author{M. Y. Yu}
\affiliation{Institute of Fusion Theory and Simulation, Zhejiang
University, Hangzhou, China, 310027}\affiliation{Institut f\"ur
Theoretische Physik I, Ruhr-Universit\"at Bochum, D-44780 Bochum, Germany}

\author{T. Tajima}
\affiliation{Fakult\"at f.\ Physik, LMU M\"unchen, Garching,
Germany, D-85748}

\author{X. Q. Yan}
\email[]{X.Yan@pku.edu.cn} \affiliation{Key Laboratory of HEDP of
the Ministry of Education, CAPT,and State Key Laboratory of Nuclear
Physics and Technology, Peking University, Beijing, China, 100871}

\author{X. T. He}
\email[]{xthe@iapcm.ac.cn} \affiliation{Key Laboratory of HEDP of the
Ministry of Education, CAPT,and State Key Laboratory of Nuclear
Physics and Technology, Peking University, Beijing, China,
100871}\affiliation{Institute of Applied Physics and Computational
Mathematics, Beijing, China, 100088}

\date{\today}

\begin{abstract}
Proton acceleration by ultra-intense laser pulse irradiating a
target with cross-section smaller than the laser spot size and
connected to a parabolic density channel is investigated. The
target splits the laser into two parallel propagating parts, which
snowplow the back-side plasma electrons along their paths,
creating two adjacent parallel wakes and an intense return current
in the gap between them. The radiation-pressure pre-accelerated
target protons trapped in the wake fields now undergo acceleration
as well as collimation by the quasistatic wake electrostatic and
magnetic fields. Particle-in-cell (PIC) simulation shows that
stable long-distance acceleration can be realized, and a 30 fs
monoenergetic ion beam of $>10$ GeV peak energy and $<2.0^\circ$
divergence can be produced by a $9.8\times10^{21}$ W/cm$^2$
circularly polarized laser pulse.

\end{abstract}
 \maketitle


Laser-driven ion acceleration has been of much recent interest
because of its broad scientific and technical applications.
Protons with tens to hundreds MeV energies are useful for cancer
therapy \cite{therapy}, high resolution imaging
\cite{imaging03,li}, fast ignition in inertial confinement fusion
\cite{fastignition}, etc., and ions with still higher energies are
relevant to high energy physics research \cite{accelerator},
laboratory modeling of astrophysical phenomena
\cite{astrophysics}, etc.

Radiation pressure acceleration (RPA) is promising for obtaining
high-quality ion beams efficiently \cite{Kli2008,Rob2008,Yan2008}.
Existing studies have shown that GeV proton beams can be obtained
by RPA using CP lasers with intensity above $10^{22}$ W/cm$^2$
\cite{xyan,qiaobin}. However, because of transverse instabilities
and hole-boring by the laser pulse \cite{xyan,qiaobin,zhang}, the
acceleration length is rather limited and it is difficult to
enhance proton energy without still higher laser intensity.
Recently, it has been shown that tens GeV proton beams
\cite{shen,yu} can be generated by a moving double-layer in a
possibly unlimited acceleration regime, or in a two-phase
acceleration regime \cite{subTeVproton,TeVcarbon} by
ultra-relativistic lasers. In classical RPA and laser wakefield
acceleration (LWFA), however, the accelerated ion bunch tends to
diverge because of the ubiquitous space-charge field
\cite{qiaobin10,bulanov10,zhangxm}. The resulting defocusing leads
to very low ion flux. Moreover, the effective laser-plasma
interaction distance is also limited by laser diffraction. As a
result, ions cannot remain collimated and be stably accelerated
over a long distance.



In this Letter, we propose a stable proton acceleration scheme
using an ultra-intense circularly polarized (CP) laser pulse and a
mass-limited target (MLT) \cite{mlt} connected to a parabolic
underdense plasma channel. The target splits the laser pulse into
two separate but adjacent parts, which are then refractively
guided by the preformed underdense plasma channel behind it. The
quasistatic longitudinal accelerating and transverse collimating
fields associated with the electron return current at the center
of the channel can stably trap, collimate, and further accelerate
the RPA target protons by LWFA over a long distance.
In contrast to the existing acceleration schemes, both proton
defocusing and laser diffraction do not occur. PIC simulations
show that ultrashort (30 fs) monoenergetic proton beams with peak
energy $>10$ GeV and divergence angle $\le 2.0^\circ$ can be
obtained with a $\sim 10^{22}$ W/cm$^2$ circularly polarized (CP)
laser pulse.

To demonstrate the proposed scheme we use the two-dimensional (2D)
PIC code KLAP \cite{xyan,zheng2002}. The CP laser pulse is of
wavelength $\lambda=1\ \mu$m and peak intensity
$I_0=9.8\times10^{21}$ W/cm$^2$ (or the normalized laser parameter
$a_0=60$). It has a super-Gaussian radial profile with spot size
$2\sigma=40\lambda$ and a $18T$ flattop envelope, where $T$ is the
laser period. The fully ionized uniform hydrogen target of density
of $N=20n_c$, where $n_c$ is the critical density, is located in
$50\lambda\leq x\leq 50.8\lambda$, $-10\lambda\leq
y\leq10\lambda$. Behind the hydrogen target is an underdense
plasma channel with a parabolic density profile $n=n_1+\Delta
n(y^2/y_0^2)$, where $n_1=0.15n_c$, $\Delta n=2n_c$,
$y_0=50\lambda$, and it is located in $50.8\lambda< x\leq
600\lambda$, $-40\lambda\leq y\leq 40\lambda$. The simulation box
is of size $100\lambda\times100\lambda$, corresponding to a moving
window consisting of $3200\times1000$ cells, with each cell
containing 800 macroparticles for the target and 6 for the
underdense channel.

\begin{figure}
\includegraphics[width=0.45\columnwidth]{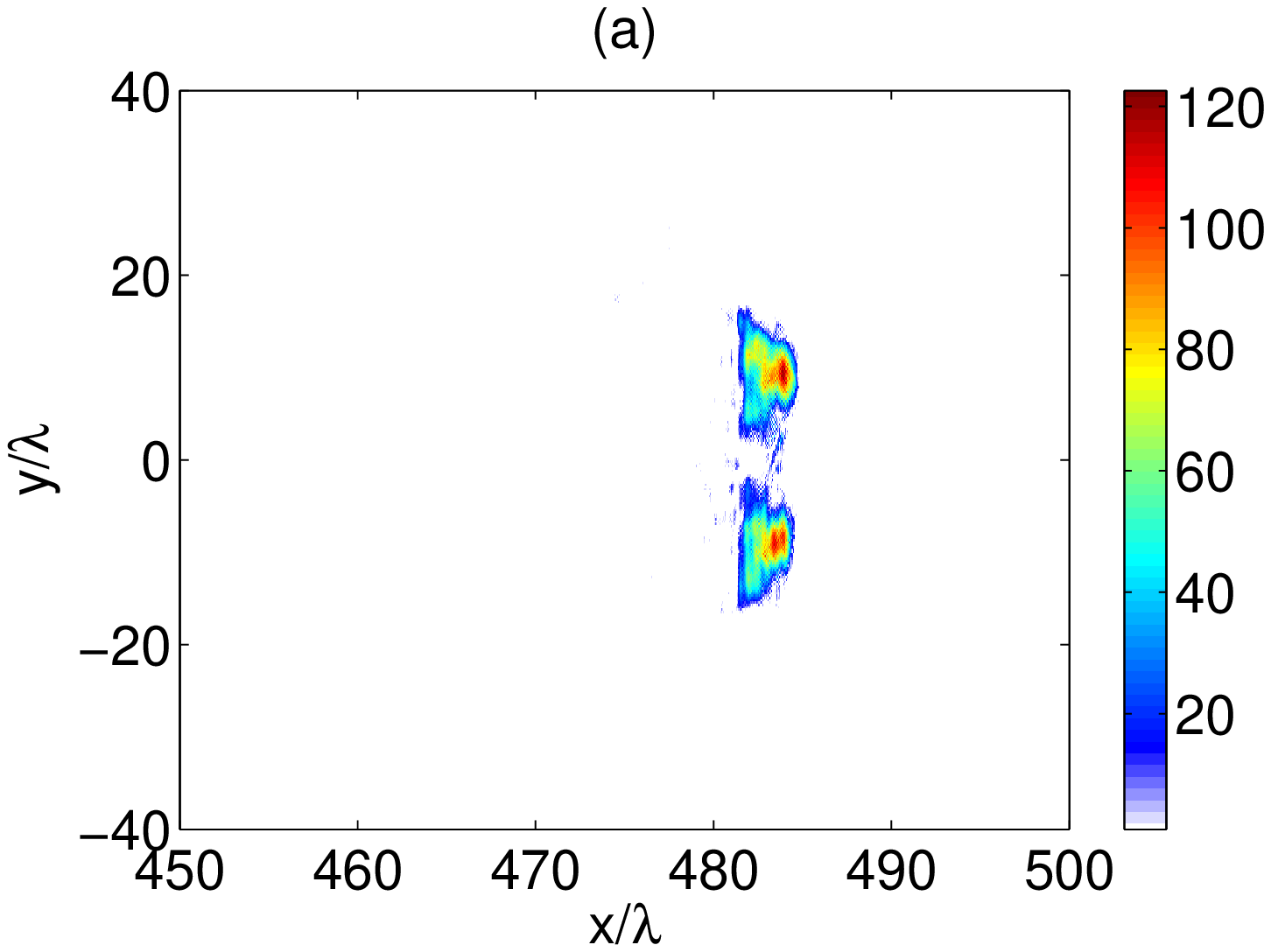}\includegraphics[width=0.45\columnwidth]{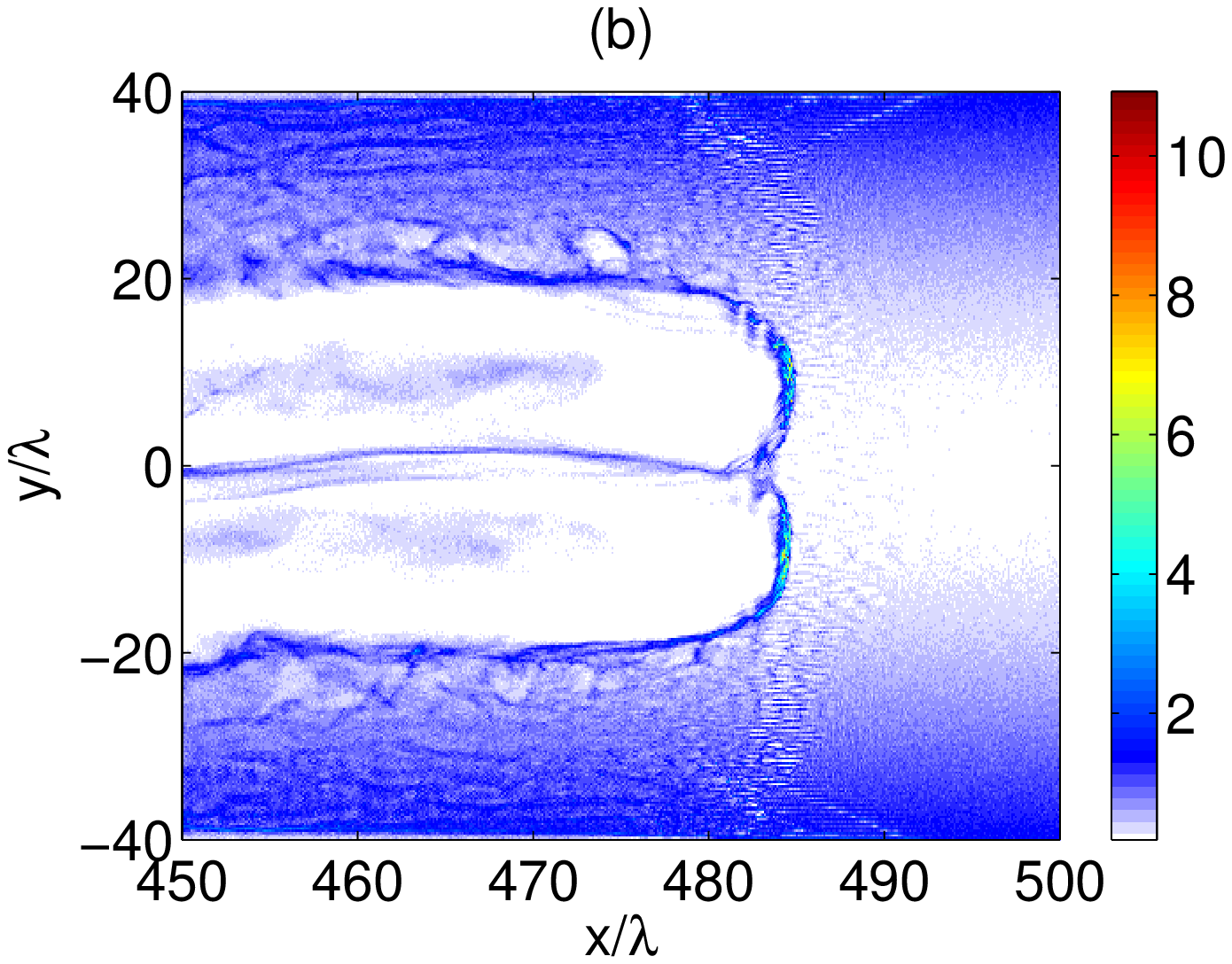}\\
\includegraphics[width=0.45\columnwidth]{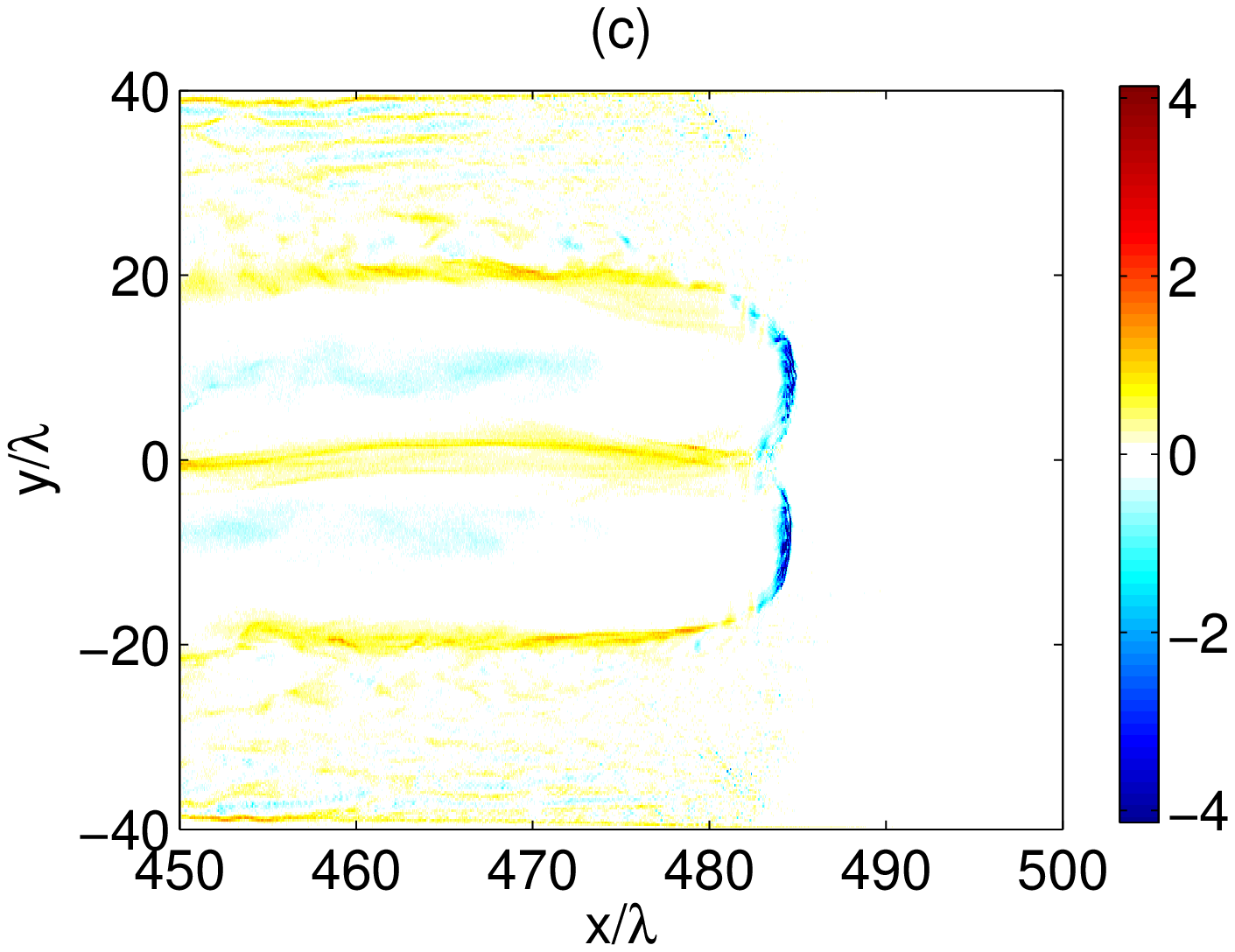}\includegraphics[width=0.45\columnwidth]{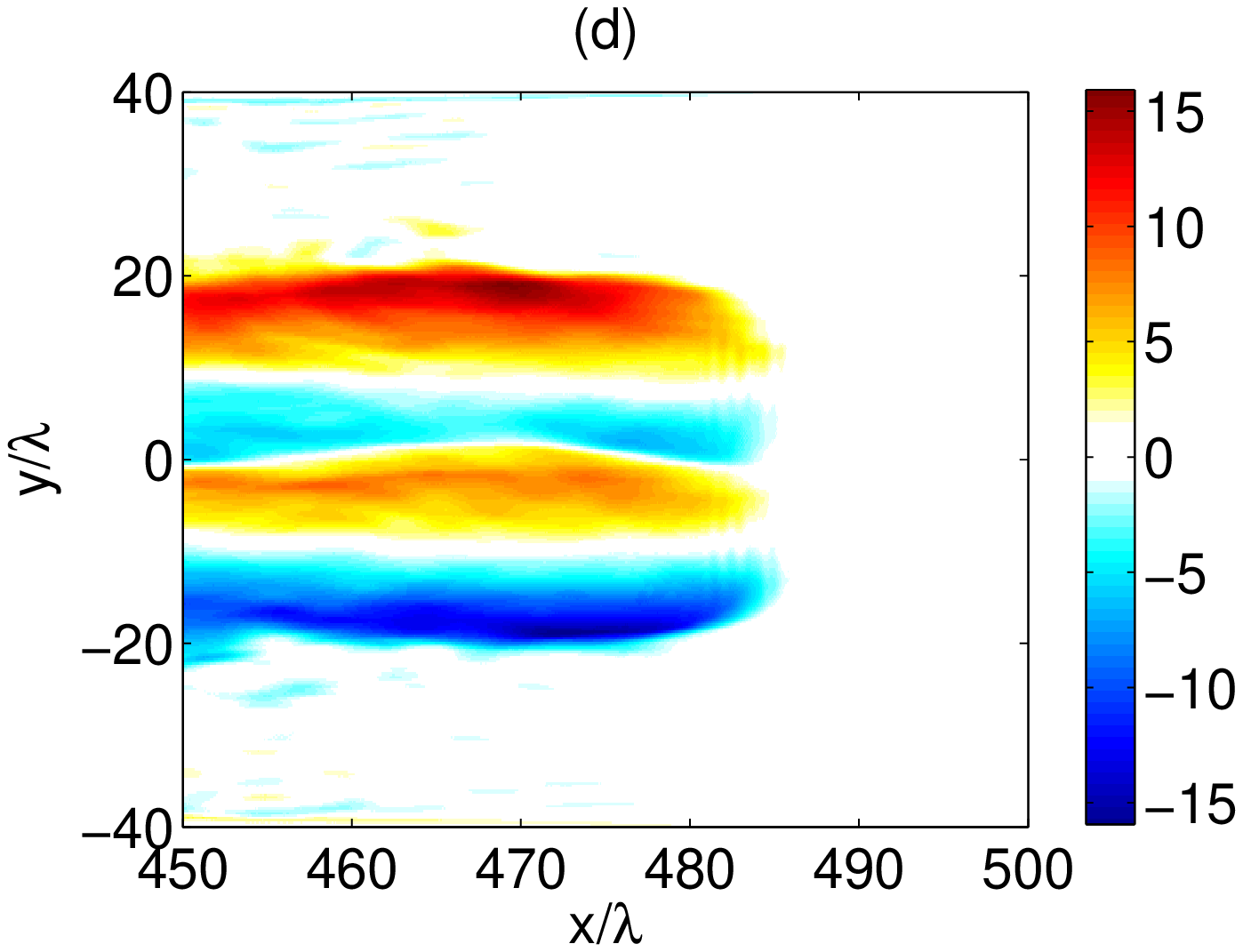}
\caption{(Color online) (a) Laser electric field
$\sqrt{E_y^2+E_z^2}$, normalized by ${\cal E}_0$. (b) Electron
density normalized by $n_c$. (c) Current density $j_x$ normalized by
$en_cc$. (d) Spatial distribution of transverse quasistatic field
$E_{ys}-B_{zs}$ (in the $\pm y$ directions), at
$t\sim500T$.\label{fig.1}}
\end{figure}

Figs.\ \ref{fig.1}(a) and (b) show the laser intensity and
electron density, respectively, at $t\sim500T$. The transverse
size $h$ of the target is smaller than the laser spot size
$\sigma$. The incoming laser pulse is split into two by the
target. After passing the latter they continue to propagate
forward in the preformed backside underdense plasma channel as
adjacent twin pulses. The ponderomotive force of the latter expels
the local plasma electrons transversely outward, creating two
adjacent wakes, or twin wakes, which trap the RPA target protons.
Fig.\ \ref{fig.1}(c) shows the longitudinal current density $j_x$
at $t\sim500T$. One can see that there is an intense electron
return-current sheet between the two wakes. Since the longitudinal
velocity $v_x$ of the trapped protons is near the light speed $c$,
the transverse field experienced by the energetic protons can be
written as ${\cal E}{\hat e}_y=E_{ys}{\hat e}_y+c^{-1}v_x{\hat
e}_xc\times B_{zs}{\hat e}_z \sim(E_{ys}-B_{zs}){\hat e}_y$, where
$E_{ys}$ and $B_{zs}$ are the self-generated electrostatic and
magnetic fields, respectively. Fig.\ \ref{fig.1}(d) shows the
total transverse quasistatic field ${\cal E}=E_{ys}-B_{zs}$
normalized by ${\cal E}_0=m_e\omega c/e$, where $m_e$ and $-e$ are
the electron mass and charge, and $\omega$ and $c$ are the laser
frequency and light speed, at $t\sim500T$. One can see that the
twin wakes act as a moving potential well for confining,
stabilizing, and further accelerating (now by LWFA) the RPA
protons for a long distance in underdense plasma. In addition,
since the target electrons also move with the twin wakes, the
expelled plasma electrons are prevented from rapidly refilling the
back side of the channel, so that the latter remains open for a
long distance. This scenario differs significantly from that of
the classical bubble/blowout acceleration regimes \cite{lpa}.

\begin{figure}
\includegraphics[width=0.45\columnwidth]{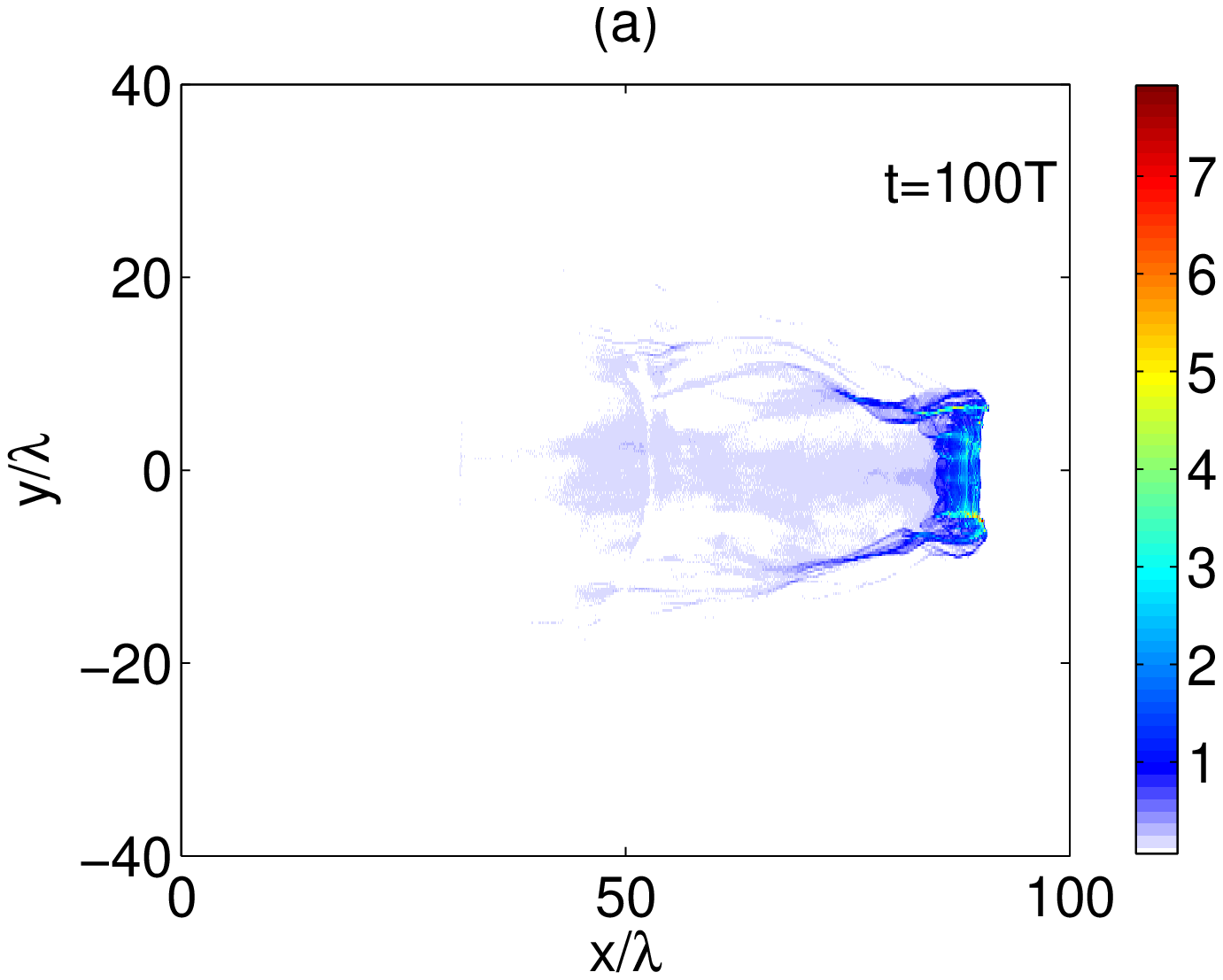}\includegraphics[width=0.45\columnwidth]{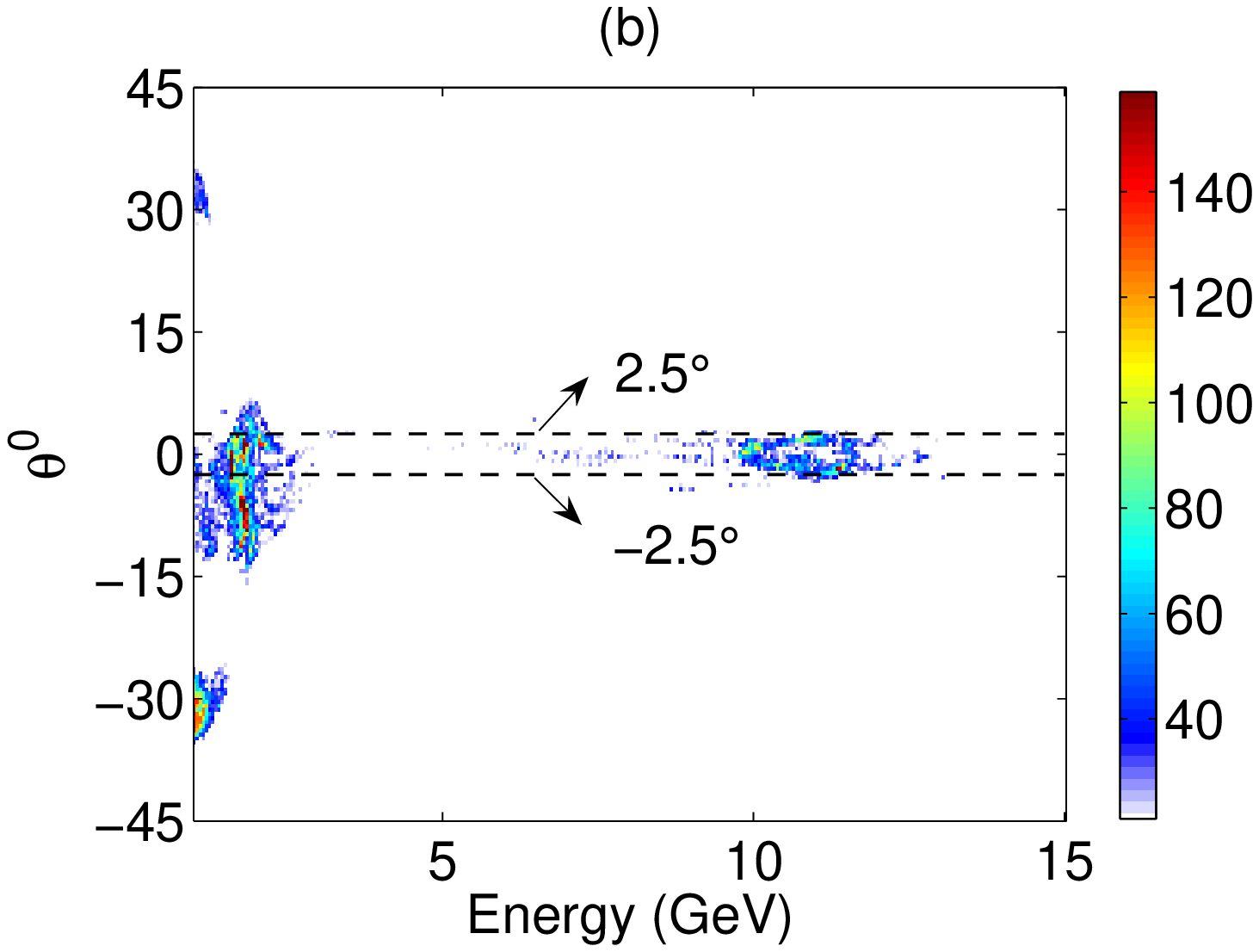}\\
\includegraphics[width=0.45\columnwidth]{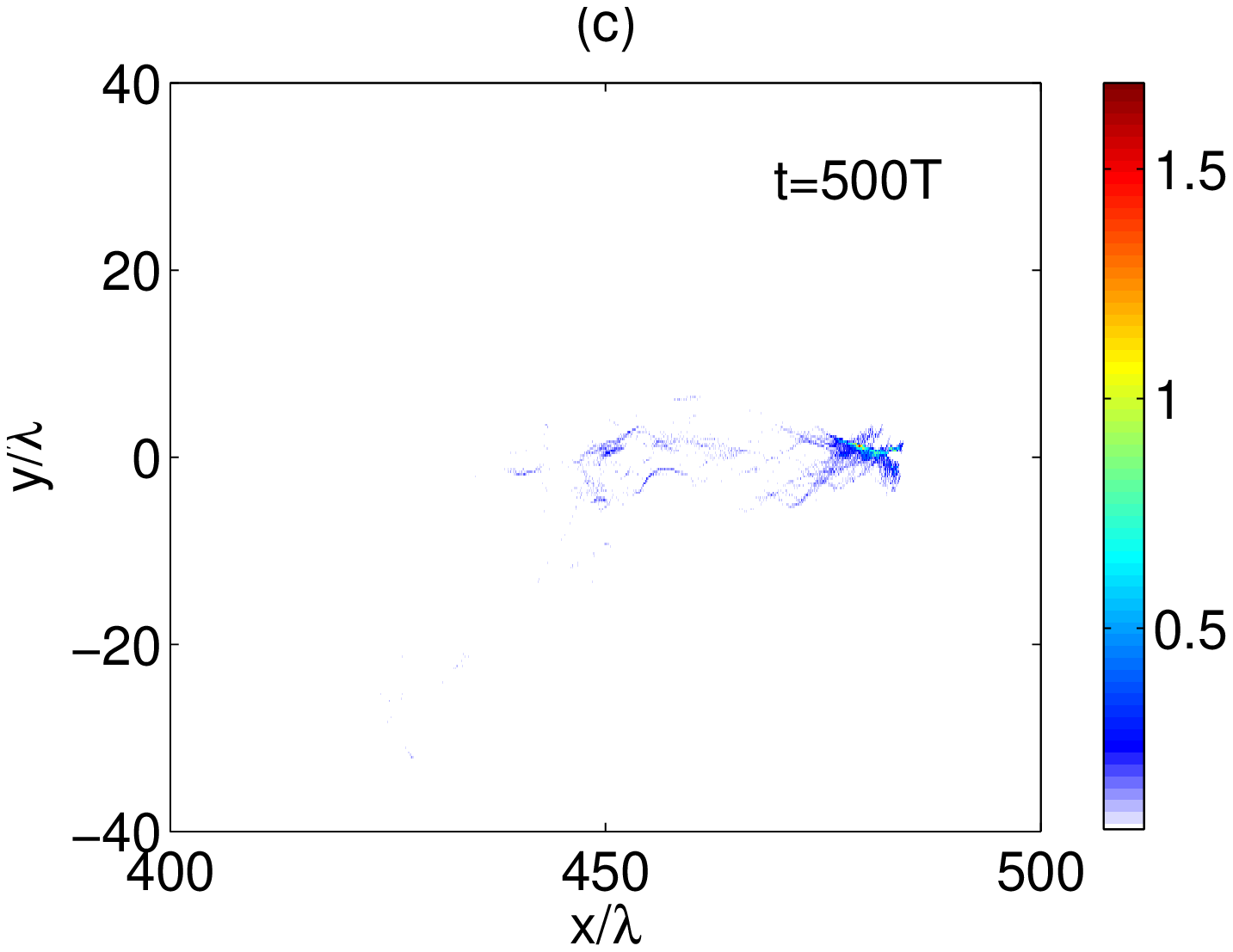}\includegraphics[width=0.45\columnwidth]{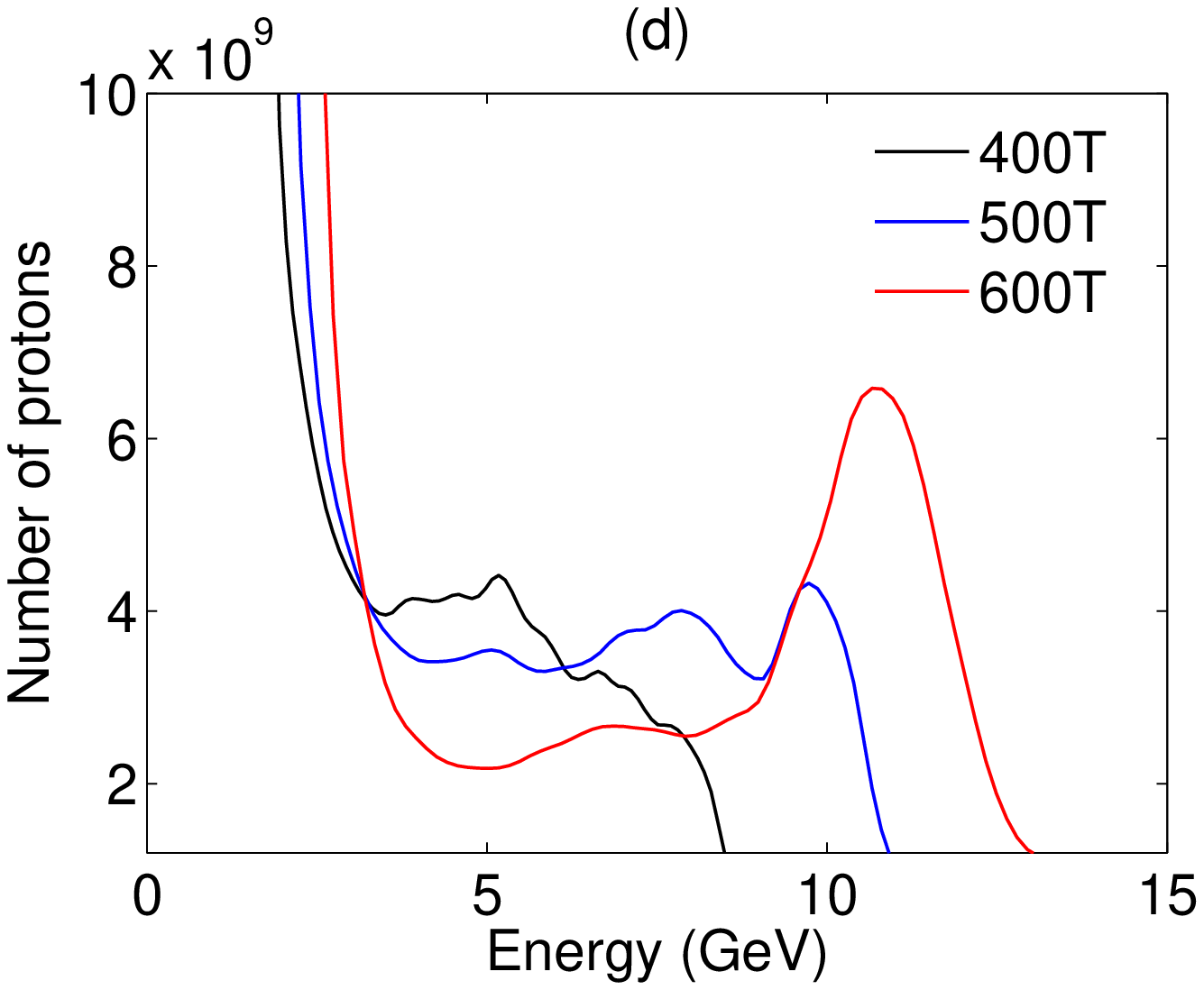}
\caption{(Color online) Density of the accelerated protons in units
of $n_c$ at (a) $t\sim100T$ and (c) $t\sim500T$. (b) Angular
distribution of the protons at $t\sim600T$. The dashed lines
correspond to the divergence angle $\pm 2.5^\circ$. (d) Evolution of
the proton energy spectrum.} \label{fig.2}
\end{figure}

When the initial laser pulse is split by the target, the effective
spot size $\sigma_1$ of each of the resulting twin light pulses is
less than that of the original pulse. Similarly, the divergence
angle $\theta_{d}=\sigma_1/Z_{R}$ is larger and the Rayleigh
length $Z_{R}=\pi\sigma_1^2/\lambda$ smaller. Earlier results have
also shown that relativistic self-focusing cannot prevent
diffraction of intense light pulses \cite{shen,yu}. Thus, without
the underdense plasma channel at the back of the target the
laser-plasma interaction distance would be very limited.

However, this undesirable scenario is completely changed by the
presence of the preformed parabolic density channel, here with the
profile $n=[n_1+\Delta n(y^2/y_0^2)]$, at the back of the target.
The dispersion relation for relativistic light waves is
$\omega^2=c^2k^2+\omega_p^2/\gamma$, where $k$ and
$\omega_p^2/\gamma$ are the wave number and effective background
plasma frequency, respectively, and $\gamma\sim\sqrt{1+a^2}$ is
the relativistic factor associated with the electron quiver
motion. For an ultra-intense CP laser pulse ($a\gg1$), the index
of refraction is
\begin{equation}\label{2}
\eta(y)\sim \left[1-\frac{1}{a(y)n_c}(n_1+\Delta n\left(\frac{y^2}{y_0^2}\right)\right]^{1/2},
\end{equation}
so that refractive light guiding (RLG) occurs if $d_y\eta<0$.
Light diffraction can then be canceled by RLG if the maximum beam
focusing angle is larger than the divergence angle $\theta_{d}$,
or
\begin{equation}
\label{3}
{\frac{n_1}{2n_c}}\left[\frac1{a(\sigma)}-\frac1{a(h/2)}\right]
+{\frac{\Delta n}{n_c}}\left[\frac{\sigma^2}{a(\sigma)y_0^2}
-\frac{(h/2)^2}{a(h/2)y_0^2}\right]\geq \left(\frac{\lambda}{\pi
\sigma_1}\right)^2,
\end{equation}
where the three terms of the inequality represent relativistic
self-focusing guiding, preformed density channel RLG, and light
diffraction, respectively. We note that this relation differs
considerably from that of
the $a\ll1$ case. 
For the parameters under consideration, the self-focusing and RLG
terms in Eq.\ (\ref{3}) have the values
$\sim0.0021$ and
$0.013$, respectively. The latter alone is already much larger
than the diffraction term, which has the value $0.0041$. One can
also see that without RLG the proton bunch would diverge. Fig.\
\ref{fig.1}(a) shows the profile of the light field at
$t\sim500T$. We see that RLG by the preformed parabolic density
channel indeed results in long distance propagation of the twin
light pulses and the corresponding twin wakes. In fact, the
laser-plasma interaction distance is effectively increased to
$\sim7Z_{R}$. Moreover, the focusing effect of RLG also doubles
the intensity of the twin pulses to $a_1=120$, as can be seen in
Fig.\ \ref{fig.1}(a).

\begin{figure}
\includegraphics[width=0.45\columnwidth]{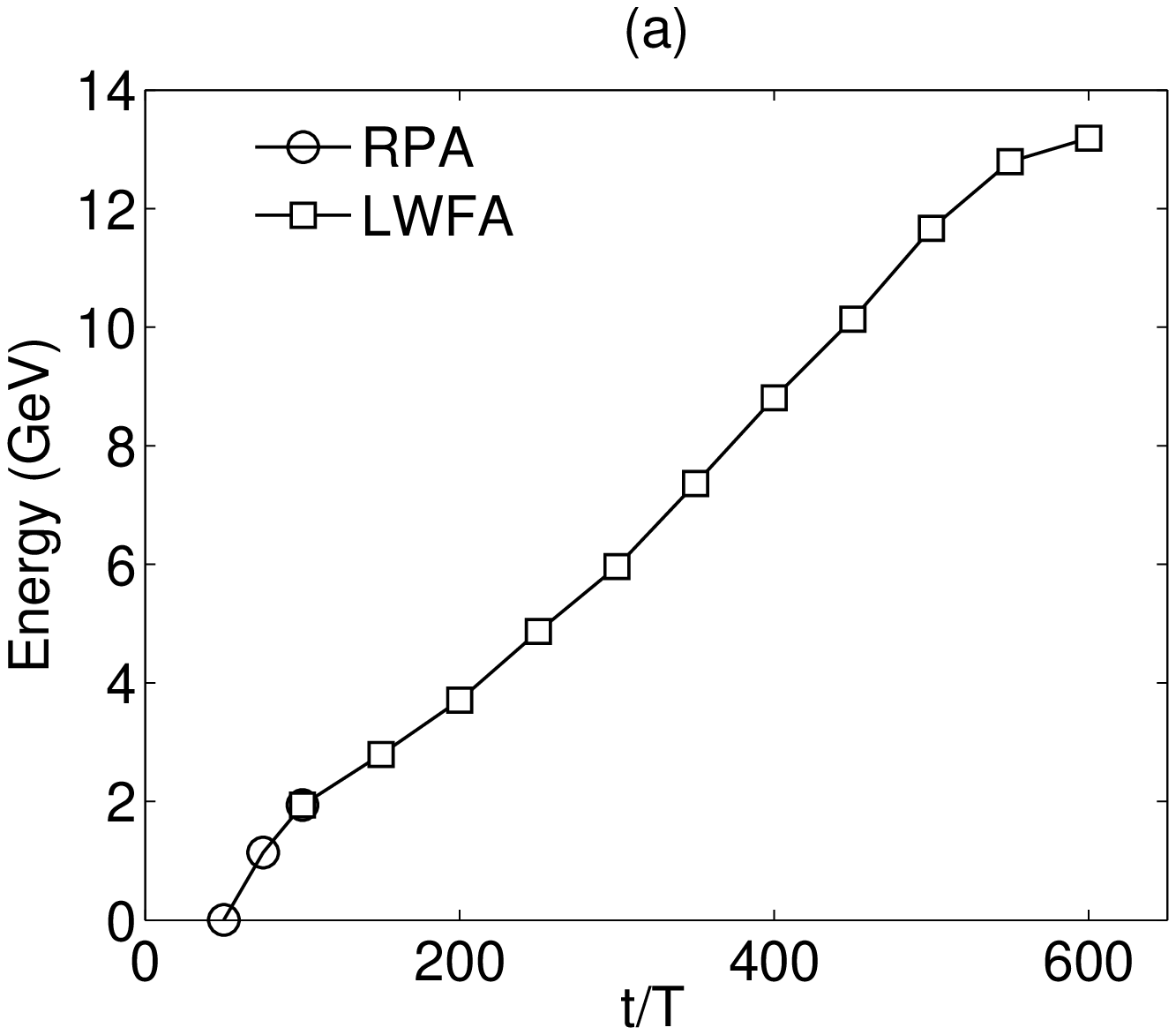}\includegraphics[width=0.45\columnwidth]{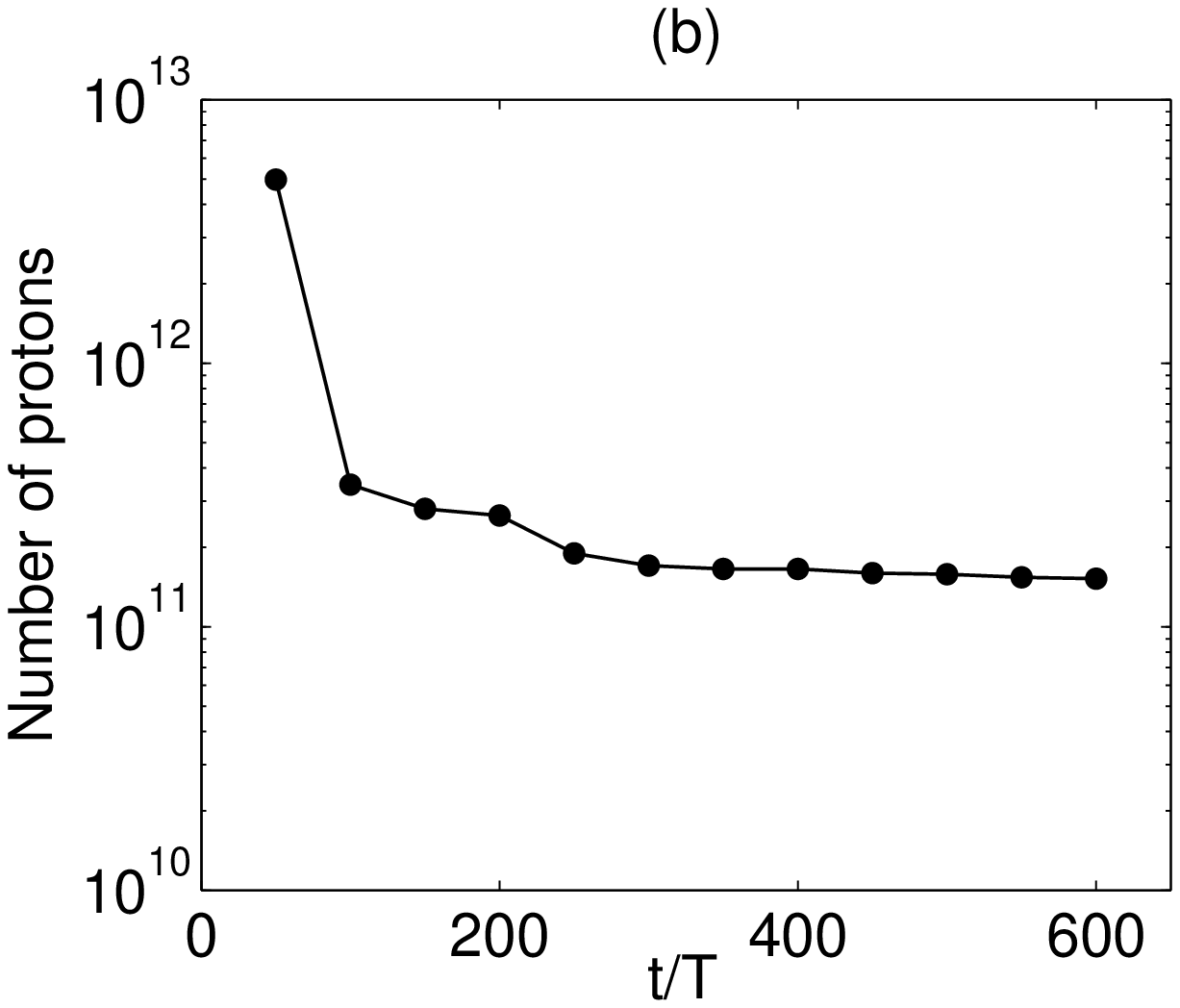}
\caption{(Color online) Evolution of the (a) maximum proton energy,
and (b) number of protons in the axis region($|y|\leq3\lambda$).
\label{fig.3}}
\end{figure}

We now consider the evolution of the pre-accelerated protons in
more detail. From Figs.\ \ref{fig.2}(a) and (c) one sees that the
relativistic protons are confined along the original laser axis
and they can propagate a long distance without divergence. Fig.\
\ref{fig.1}(d) shows that the maximum quasistatic transverse (or
focusing) field ${\cal E}_{\max}$ induced by the charge separation
and the return current at the center of the channel is about
$6{\cal E}_0$, agreeing well with the theoretical prediction
${\cal E}_{\max}/{\cal E}_0
\sim(\pi/2)({a_1n_1}/n_c)^{1/2}$\cite{lpa}. 
The angular distribution of the accelerated protons at $t\sim600T$
is given in Fig.\ \ref{fig.2}(b), which shows that the average
emission angle for protons with energy greater than 8 GeV is less
than $2.0^\circ$. That is, the output proton beam is well
collimated by the transverse quasistatic field.

Fig.\ \ref{fig.3}(a) shows the evolution of the maximum proton
energy. One can see the two stages of the acceleration process,
namely RPA and LWFA. In the initial RPA regime, the maximum proton
velocity reaches $0.945c$, or 1.93 GeV, at $t\sim 100T$. The
pre-accelerated protons then undergo LWFA in the twin wakes until
the light pulses are almost depleted at $t\sim 600 T$. At that time,
the trapped proton bunch approaches the front of the twin wakes,
indicating that the dephasing length is roughly the same as the pump
depletion length. We also see that the maximum proton energy
increases rapidly with time and then saturates at $t\sim600 T$.
Fig.\ \ref{fig.2}(d) shows the evolution of energy spectrum. One can
see that the latter improves with time and eventually achieves a
sharp peak at about 12 GeV, which is also consistent with that of a
theoretical model \cite{subTeVproton} $W_{\max}\simeq
(a_1^2n_c/n_1)m_ec^2/6$. Fig.\ \ref{fig.3}(b) shows that the number
of protons in the region $|y|\leq 3\lambda$ first decreases and then
becomes constant at $1.5\times10^{11}$, indicating that the
transverse spreading effect for the energetic protons is well
controlled in the backside channel. Their total energy is above
$500$ J and the laser conversion efficiency is about $10\%$. The
accelerated protons are well compressed in the phase space and the
quasi-monoenergetic pulse has a duration of about $30$ fs. Such a
short proton pulse can be used to excite a plasma wakefield that can
in turn accelerate the plasma electrons \cite{allen}.


\begin{figure}
\includegraphics[width=0.45\columnwidth]{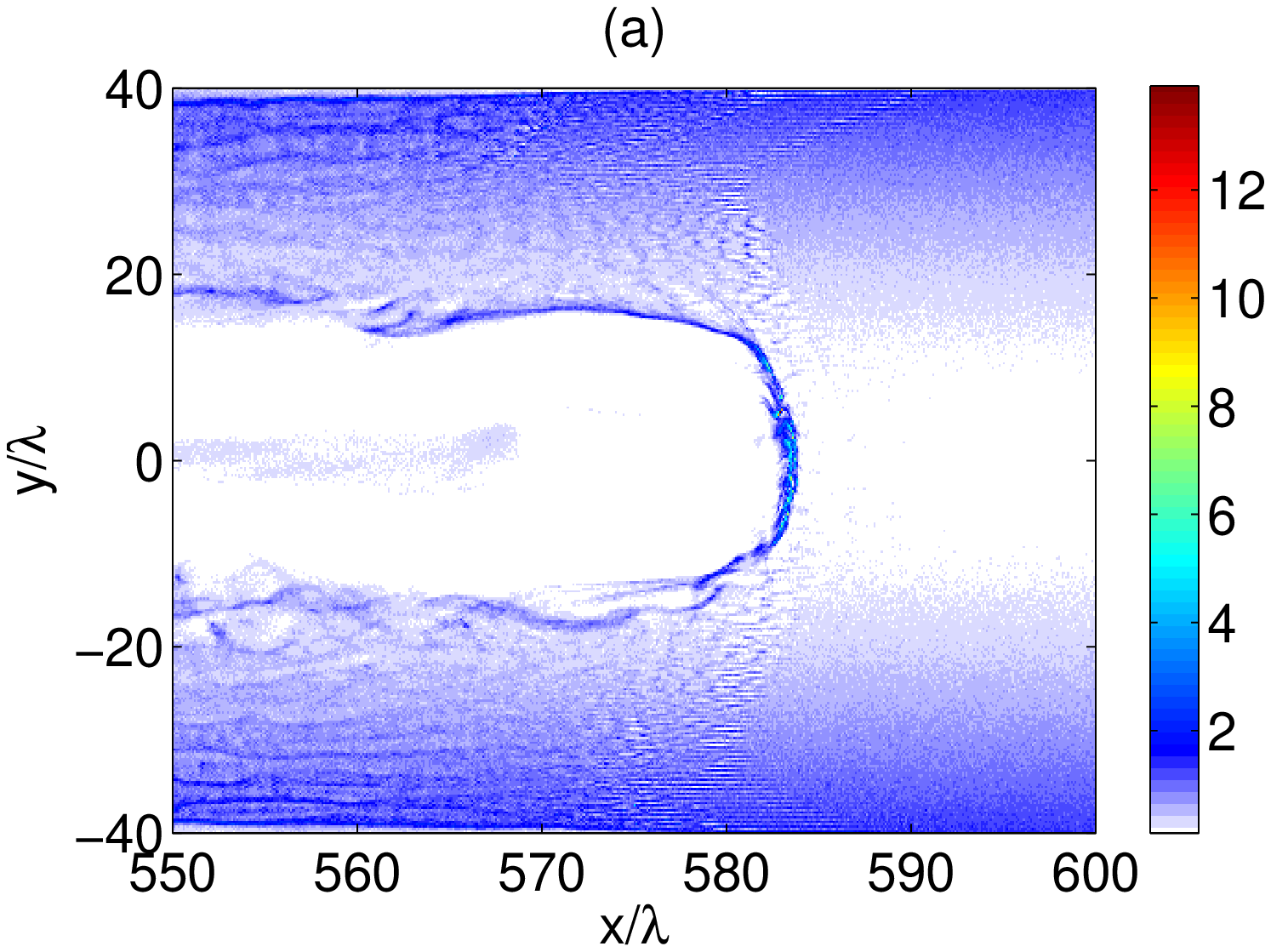}\includegraphics[width=0.45\columnwidth]{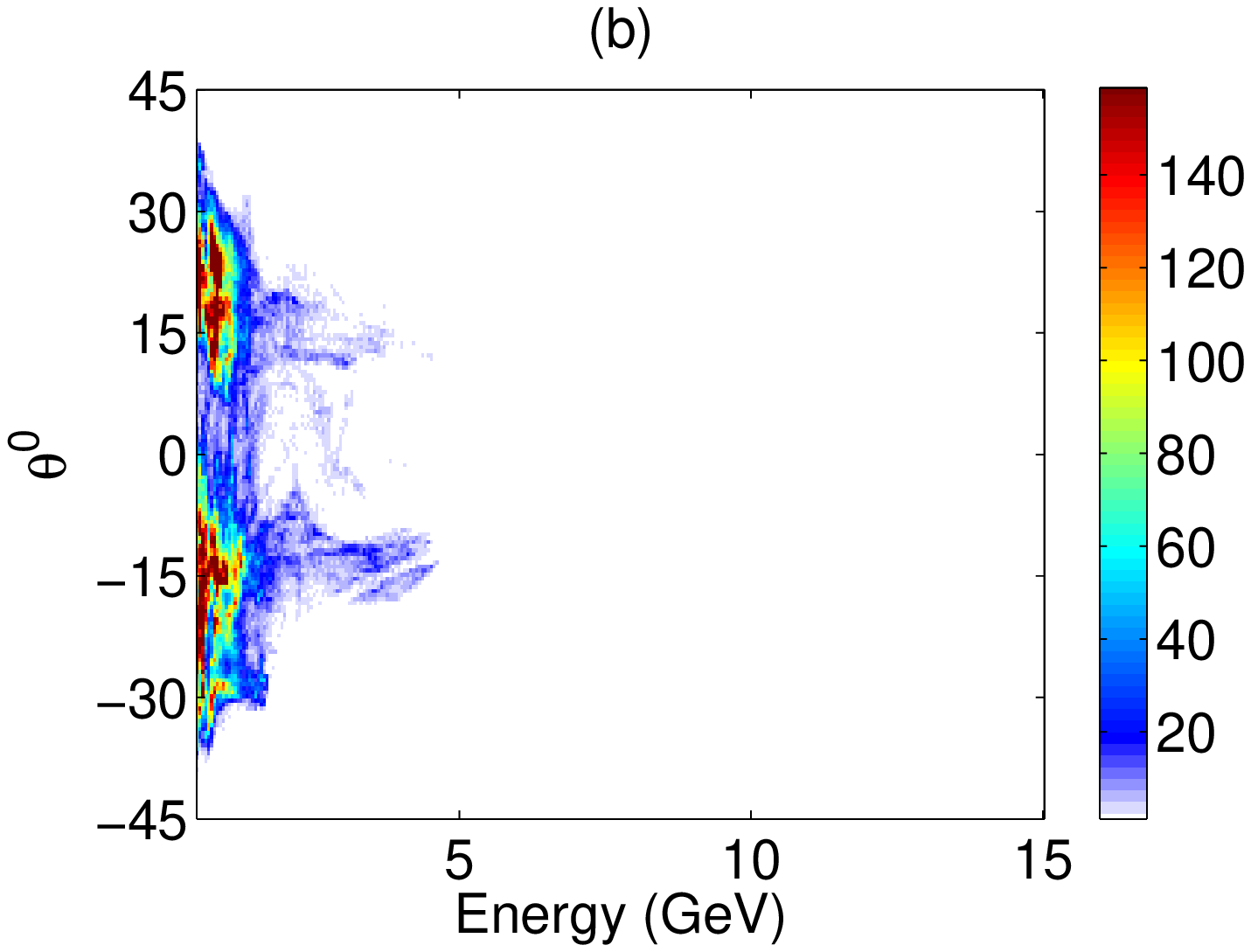}\\
\includegraphics[width=0.45\columnwidth]{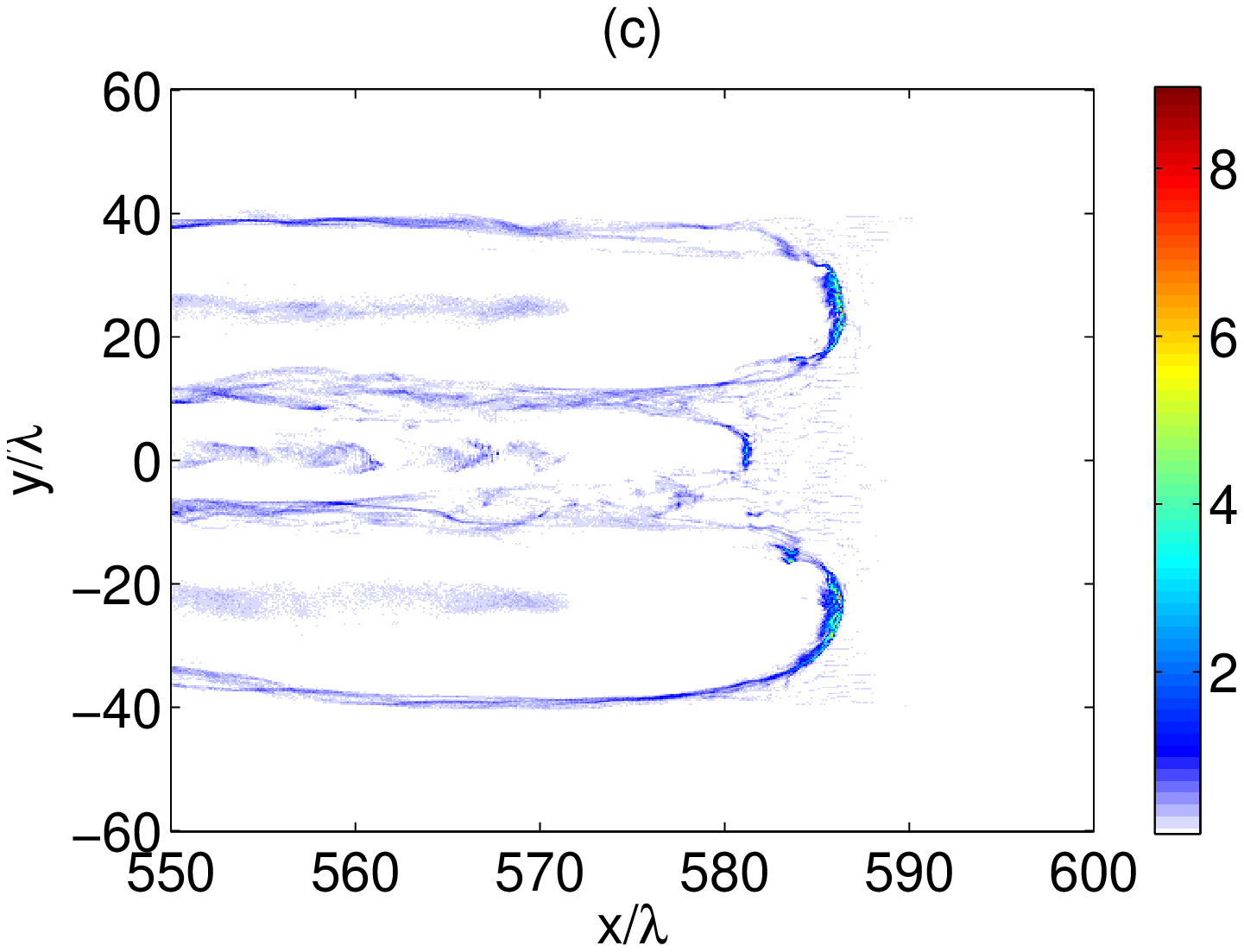}\includegraphics[width=0.45\columnwidth]{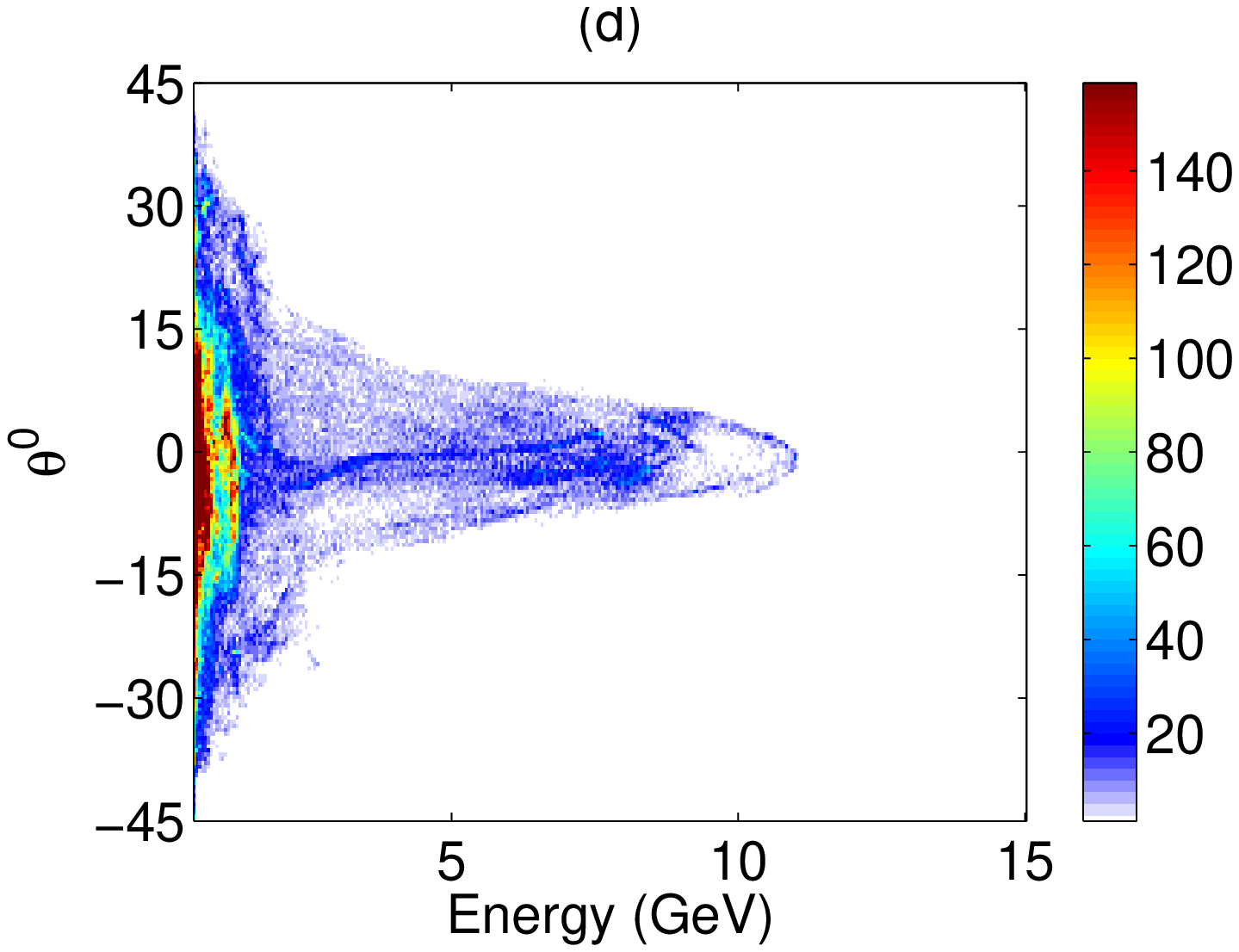}
\caption{(Color online) Electron charge distributions at $t\sim600T$
for the (a) single-wake (when the condition $h\le\sigma$ is not
satisfied) and (c) twin-wakes cases. The corresponding angular
distributions of the protons are shown in (b) and (d),
respectively.\label{fig.4}}
\end{figure}

A key issue in the proposed scheme is the formation of a stable
proton focusing and collimating twin-wakes structure that moves at
relativistic speed with the target-modulated laser light. When the
transverse size of target is too large comparing to the laser spot
size, the initial laser pulse can fully interact with the target.
The laser is not split into two parts and no twin wakes are
formed. Without the transverse confinement in the underdense
region, the RPA protons would diverge. This can be seen in the
simulation by setting $h=2\sigma=40 \mu$m, with the other laser
and plasma parameters the same as in the preceding simulations.
Fig.\ \ref{fig.4}(a) shows that instead of the stable twin wakes
(see Fig.\ \ref{fig.1}(b)) a single channel is generated by the
laser pulse. Due to the defocusing effect of the space-charge
field, the RPA protons quickly diverge in the wake, as can be seen
in Fig.\ \ref{fig.4}(b) for the emission angle (almost $20^\circ$)
at $t\sim600 T$. In addition, protons that deviate from the axis
region do not experience LWFA. The maximum energy of the
accelerated protons is about 5 GeV, which is lower than that in
the the stable twin-wakes case (see Fig.\ \ref{fig.2}(d)). The
quality of the final proton bunch is thus relatively poor in terms
of energy gain as well as emission angle.

The condition (\ref{3}) for guiding the laser light in the
underdense plasma is also important. If the maximum focusing angle
is less than the beam divergence angle, RLG of the twin pulses in
the density channel cannot occur. Instead, the acceleration
process becomes unstable \cite{shen,yu}. To see this, we set
$\Delta n=0$ (uniform plasma behind the target), with the other
parameters unchanged. That is, the density-channel guiding effect
is removed. Since the relativistic self-focusing term is less than
the beam diffraction term. Fig.\ \ref{fig.4}(c) shows the density
distribution of electrons in this case. We see that although a
twin-wakes structure is still formed at the beginning, it does not
survive for a long time. Only a fraction of the RPA protons is
further accelerated and for a much shorter distance than that in
the preformed channel. In addition, the average divergence angle
is now rather large, namely $14^\circ$, as shown in Fig.\
\ref{fig.4}(d).


In conclusion, an efficient laser-driven proton acceleration scheme
using a small target with a preformed parabolic underdense channel
behind is proposed. The laser pulse drives RPA protons and is split
by the small target into two parts, which excite behind them a
twin-wakes structure in the channel. The pre-accelerated RPA target
protons are efficiently collimated and further accelerated by LWFA
in the wake. With this scheme, both light diffraction and proton
defocusing are overcome, making long-distance acceleration possible.
The conditions and characteristics of proton acceleration from the
PIC simulations agree well with that from the analytical estimates.
It is shown that ultrashort ($30$ fs) highly-collimated
monoenergetic proton bunches with peak energy $>10$ GeV can be
produced with a $60$ fs CP laser pulse of intensity $9.8\times
10^{21}$ W/cm$^2$. These bunches may be superior in fs metrology in
determining the validity of the OPERA type experiments for neutrino
velocity \cite{netrino}. Such a short and collimated proton pulse
can also be promising to excite a plasma wakefield for TeV electron
acceleration.


\begin{acknowledgments}
This work is supported by the National Natural Science Foundation of
China (Grant nos. 10935002, 10974022,11025523,11175026, 11175029),
and the National Basic Research Program of China (Grant nos.
2008CB717806, 2012CB801111).
\end{acknowledgments}

\bf
\end{document}